\documentclass{aa}
\usepackage{graphicx}
\usepackage{txfonts}

\def\<#1>{\langle\hbox{$#1$}\rangle}
\def\hoverarrow#1{\setbox0\hbox to 0pt{\hss$\scriptstyle\rightarrow$}%
                 #1\kern.6ex\raise 1.6ex\box0\kern-.2ex}
\def\btheta{{\hoverarrow\theta}}
\def\bbeta{{\hoverarrow\beta}}
\def\witchbox#1#2#3{\hbox{$\mathchar"#1#2#3$}}
\def\leqsim{\mathrel{\rlap{\lower3pt\witchbox218}\raise2pt\witchbox13C}}
\def\zl{z_{\rm L}}  \def\zs{z_{\rm S}}
\def\Dl{D_{\rm L}} \def\Ds{D_{\rm S}} \def\Dls{D_{\rm LS}}
\def\thee{\theta_{\rm E}}
\def\Tiso{T_{\rm iso}}
\def\Gyr{\,{\rm Gyr}}
\def\fr#1#2{\hbox{$#1\over#2$}}
\def\0{\phantom0}

\begin{document}

\title{Time-delay quasars: scales and orders of magnitudes}

\author{Prasenjit Saha\inst{1,2}}

\institute{Astronomy Unit, 
           Queen Mary and Westfield College,
           University of London,
           London E1~4NS, UK \\
           \email{p.saha@qmul.ac.uk}
           \and
           Observatoire astronomique,
           11, rue de l'Universit\'e,
           Strasbourg F-67000, France}

   \date{Received 12 August 2003/Accepted 18 September 2003}

\abstract{We can think of a lensed quasar as taking the Hubble time,
shrinking it by $\sim10^{-11}$, and then presenting the result to us
as a time delay; the shrinking factor is of the order of fractional
sky-area that the lens occupies.  This cute fact is a straightforward
consequence of lensing theory, and enables a simple rescaling of time
delays.  Observed time delays have a 40-fold range, but after
rescaling the range reduces to 5-fold.  The latter range depends on
details of the lens and lensing configuration---for example, quads
have systematically shorter rescaled time delays than doubles---and is
as expected from a simple model.  The hypothesis that observed
time-delay lenses all come from a generalized-isothermal family can be
ruled out.  But there is no indication of drastically different
populations either.
}

\maketitle

\section{Introduction}

Most of the observables in gravitational lensing (image positions and
magnifications) are intrinsically dimensionless.  The exception is the
time delay between images, which takes its dimensionality straight
from the universe:\footnote{This point appears to have been first
emphasized by \cite{n90}, although it is implicit already in
\cite{r64}.}  $\Delta t \propto H_0^{-1}$.  This remarkable fact is
the essential reason for much research effort going into measuring
time delays.  The observations have been increasingly successful---in
1995 there was but one controversial time delay, currently there are
nine non-controversial ones. These are summarized in Table~\ref{tabby}
below.

But curiously, even as the image and time delay data have improved,
the error bars on the inferred $H_0$ have not.  As an example,
consider 0957+561.  Between \cite{kundic97} and \cite{oscoz01} the
time-delay value changed by only 2\%.  But meanwhile, whereas
\cite{kundic97} quote $H_0=64\pm13$ (95\% confidence) in the usual
units of $\rm km\,s^{-1}\,Mpc^{-1}$, \cite{bernstein99} with more
imaging and more modelling conclude that the data imply only
$77^{+29}_{-24}$, while \cite{keeton00} assert that further data on
the lensed host galaxy invalidates all previously published models,
and they decline to give an $H_0$ estimate at all.  Basically, the
problem is that simple lens models are unable to fit the images to the
mas-level demanded by current data, while more complicated models can
fit the data but are non-unique and can produce identical observables
from very different values of $H_0$.

Modellers have responded to this dilemma with two strategies.  One is
to try to identify simple models that both have enough parameters to
fit or nearly fit the data and can be justified on galactic-structure
grounds; \cite{kochanek03} is typical of these.  The other strategy is
to try to explore the space of all plausible models allowed by the
data; \cite{rsw03} is a recent example.  For a review by authors
representing different points of view see \cite{courbin03}.

In the current context of good data and active modelling but no
consensus on models, it is interesting to step back and pose some
questions that tend to get obscured in the details of modelling.
First, we can think of the purpose of modelling time-delay lenses as
being to discover one dimensionless number, the factor relating
$\Delta t$ and $H_0^{-1}$.  What contributions to this number are
well-constrained and what are poorly constrained?  What range of
values do the data imply for the poorly-constrained part?  Is that
range systematically different for doubles and quads, and/or for
isolated lensing galaxies versus interacting galaxies?  And is that
range consistent with what we expect from popular models?  Nine
systems is a small sample, but it is enough to provide preliminary
answers to these questions, and to do so is the aim of this paper.

\section{A scaling relation for time delays}

In lensing theory the arrival time can be written as
\begin{equation}
t(\btheta) = (1+\zl){\Dl\Ds\over c\Dls}
\left[\fr12|\btheta-\bbeta|^2 - \psi(\btheta)\right]
\end{equation}
where the symbols have their usual meanings.  For convenience, let us
abbreviate this expression.  First, we write $\tau(\btheta)$ for the
expression inside square brackets.  The factor outside square brackets
equals $H_0^{-1}$ times a dimensionless distance factor $D$ (say)
that depends on redshifts and (weakly) on cosmology, but not on $H_0$;
for small $\zl$ and large $\zs$, $D\simeq\zl(1+\zl)$.  We note further
that only differences in arrival-time between images are observable.
Hence observable time delays have the form
\begin{equation}
\Delta t = H_0^{-1} \, D \, \Delta\tau
\label{dt}
\end{equation}

We expect that $\Delta\tau$ will be of the same order as
$|\btheta-\bbeta|^2$ but a few times smaller, the precise value
depending on details of lens and lens configuration. For an observed lens
we might predict
\begin{equation}
\Delta\tau \sim (\theta_1+\theta_2)^2
\end{equation}
where $\theta_1,\theta_2$ are the $\theta$ values of the first and
last images to arrive.  To focus attention on the proportionality
factor, I propose to consider the dimensionless quantity
\begin{equation}
\varphi \equiv {\Delta\tau \over \fr1{16}(\theta_1+\theta_2)^2} .
\label{skyf}
\end{equation}
We can calculate $\varphi$ from a lens model, but not directly from
observations.  We can, however, measure a related quantity, a scaled
time delay
\begin{equation}
\Delta T \equiv {\Delta t \over \fr1{16}(\theta_1+\theta_2)^2 \, D}
\label{sdel}
\end{equation}
directly from observations, and substituting equations (\ref{dt}) and
(\ref{skyf}) we see that
\begin{equation}
\Delta T = \varphi H_0^{-1} .
\end{equation}

The factor $\fr1{16}$ is ad hoc, but it allows the following
interpretation.  Recall that the image separation in a galaxy lens is
about twice the Einstein radius:
\begin{equation}
\theta_1+\theta_2 \simeq 2\thee .
\label{thee}
\end{equation}
For an isothermal, the relation (\ref{thee}) is exact.  But even for
non-circular lenses, where $\thee$ is not strictly defined, the image
configuration can be used to define an effective $\thee$.  Using
(\ref{thee}) the denominator in (\ref{skyf}) is $\pi\thee^2/(4\pi)$,
i.e., the area of the Einstein ring as a fraction of the sky.  In
other words, if we scale the observed time delay by the lens's
covering factor on the sky we get $H_0^{-1}$ times a `fudge factor' of
the order of unity.

For isothermal lenses, $\varphi$ ranges from 0 to 8, averaging
$\fr{16}3$. To see this, recall that for isothermals,
$\Delta\tau=2\thee\beta$ and note that $\beta$ could be anywhere in
the Einstein ring. Hence $\<\beta>=\fr23\thee$ and using (\ref{thee})
gives
\begin{equation}
\<\Delta\tau>_{\rm iso} = \fr13(\theta_1+\theta_2)^2 .
\label{tauiso}
\end{equation}
Equation (\ref{tauiso}) is interesting for comparison with
non-isothermals, but for isothermals themselves, we can do better.
Combining $\Delta\tau=2\thee\beta$ with $\theta_1-\theta_2=2\beta$,
which isothermals also satisfy, allows us to define
\begin{equation}
\Delta \Tiso \equiv {\Delta t \over \frac12(\theta_1^2-\theta_2^2) \, D}
\label{wmk}
\end{equation}
which equals $H_0^{-1}$.

\leavevmode \cite{wmk00} show that $\Delta \Tiso=H_0^{-1}$ is not
restricted to isothermals but is valid for a large family of
generalized-isothermal lenses, and argue that it will be generally
applicable in nature.  If so, $\varphi$ could be eliminated
altogether.  We can readily test if this is the case.

\section{Scaling the data}

We now present the obvious comparison of the scaled time delays $\Delta T =
\varphi H_0^{-1}$ with current data.

Table \ref{tabby} lists the relevant quantities for the various
time-delay systems.  The time delays references are given in the
table, and the other data are taken from the CASTLES survey and
compilation by \cite{castles}.  For quads, only the first and last
images (that is, the longest time delay) are considered, to enable a
simple comparison with doubles.  There are some caveats to the values
of $\theta_1$ and $\theta_2$: for 1830 and 0218 the lens-centre is
very uncertain and hence $\theta_1,\theta_2$ are especially uncertain,
for 1608 the lens is apparently an interacting pair of galaxies, and
0957 and 0911 are in clusters and hence have large
lensing contributions from other galaxies.

\begin{table}
\caption[]{Summary of time-delay data.}
\label{tabby}
$$
\begin{array}{p{2cm}cccl}
\hline
\noalign{\smallskip}
Object & \theta_1 & \theta_2 & D & \hfil\Delta t \\
\noalign{\smallskip}
\hline
\noalign{\smallskip}
0957+561  &\quad 5.23 \quad& 1.03 &\quad 0.49 \quad&  423\pm1^{\mathrm a} \\
0911+055  &      2.24      & 0.82 &      1.12      &  146\pm8^{\mathrm b} \\
1520+530  &      1.21      & 0.39 &      1.21      &  130\pm3^{\mathrm c} \\
2149--275 &      1.37      & 0.33 &      0.67      &  103\pm12^{\mathrm d} \\
1608+656  &      1.53      & 0.58 &      1.20      & \077\pm3^{\mathrm e} \\
1600+434  &      1.02      & 0.39 &      0.59      & \051\pm4^{\mathrm f} \\
1830--211 &      0.68      & 0.30 &      1.47      & \026_{-4}^{+5}%
                                                     {}^{\mathrm g} \\
1115+080  &      1.37      & 0.95 &      0.39      & \025\pm4%
                                                       ^{\mathrm {h,i}} \\
0218+357  &      0.22      & 0.14 &      2.42      & \010\pm1^{\mathrm j,k} \\
\noalign{\smallskip}
\hline
\noalign{\smallskip}
\end{array}
$$
{\rightskip 0pt plus 2cm
$^{\mathrm a}$\cite{oscoz01} \quad
$^{\mathrm b}$\cite{hjorth02} \quad
$^{\mathrm c}$\cite{burud02b} \quad
$^{\mathrm d}$\cite{burud02a} \quad
$^{\mathrm e}$\cite{fassnacht02} \quad
$^{\mathrm f}$\cite{burud00} \quad
$^{\mathrm g}$\cite{lovell98} \quad
$^{\mathrm h}$\cite{schechter97} \quad
$^{\mathrm i}$\cite{barkana97} \quad
$^{\mathrm j}$\cite{biggs99} \quad
$^{\mathrm k}$\cite{cohen00} \par}
\end{table}

\begin{figure}
\centering \includegraphics[width=.4\textwidth]{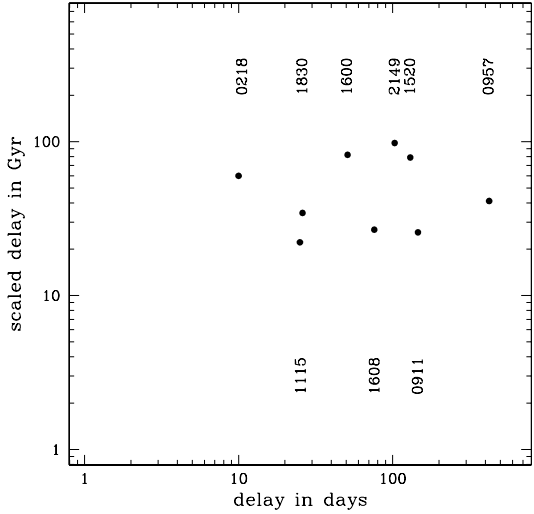}
\caption{Plot of the scaled time delay $\Delta T$ defined in
Equation~(\ref{sdel}) against the observed time delay. The various
lenses are labelled by their short names: quads are labelled below,
doubles above.}
\label{fig-skyf}
\end{figure}

\begin{figure}
\centering
\includegraphics[width=.4\textwidth]{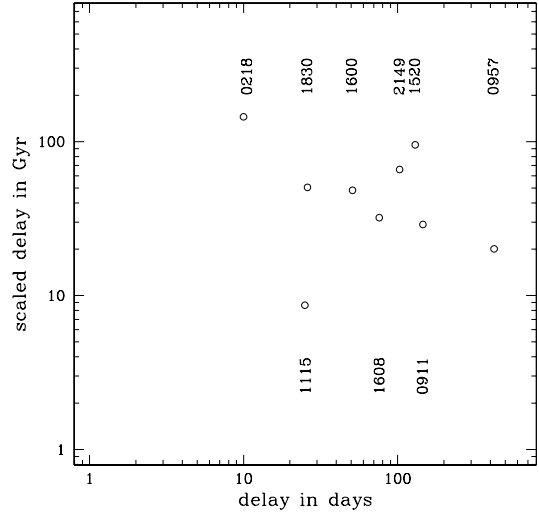}
\caption{As in Figure \ref{fig-skyf}, but omitting the $D$ factor in
the scaled time delay.}
\label{fig-simp}
\end{figure}

Figure \ref{fig-skyf} shows $\Delta T$ against $\Delta t$ for the
currently known time-delay systems.  Since error bars on time delays
are typically a few percent they are not shown here.  We notice three
things:
\begin{itemize}
\item Whereas $\Delta t$ ranges over a factor of 40, $\Delta T$ ranges
over a factor of 5.
\item No correlation is evident between $\Delta T$ and $\Delta t$.
According to the shuffling test described in Appendix A, the trend is
significant at the 75\% level---i.e., not significant.  (Meanwhile,
Figure \ref{fig-simp} shows how Figure \ref{fig-skyf} changes if we
ignore all redshift information and simply set $D=1$.  The scatter
increases, but again there is no significant trend.)
\item If we assume that $H_0^{-1}$ is $\sim15\Gyr$, then the range of
$\varphi$ is 1.5--2 for quads and about 2--6 for doubles.
[R. Ibata (personal communication) drew attention to this separation
from an early version of Figure \ref{fig-skyf}.]
\end{itemize}
The various caveats above do not appear to affect these points.

\begin{figure}
\centering
\includegraphics[width=.4\textwidth]{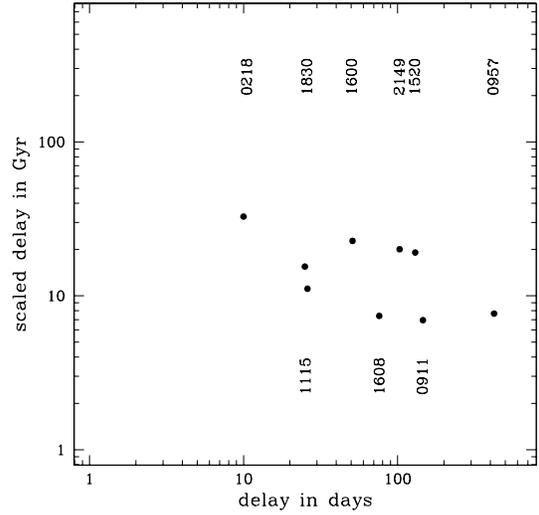}
\caption{$\Tiso$ as defined in Equation~(\ref{wmk}) against
the observed time delay. The non-physical trend is significant (see
text), and hence the generalized isothermal models are rejected.}
\label{fig-wmk}
\end{figure}

We can also compare $\Delta \Tiso=H_0^{-1}$ against the data to test
whether the lenses belong to the generalized isothermal family studied
by WMK.  Figure
\ref{fig-wmk} shows $\Delta \Tiso$ against $\Delta t$ for the same
systems.  We notice the following
\begin{itemize}
\item $\Delta \Tiso$ (expected to be constant, since there is no
$\varphi$ factor) ranges over a factor of 5.
\item Larger lenses tend to give lower $\Delta \Tiso$, and according
to the shuffling test, this non-physical trend is significant at the
95\% level.
\item Of the nine lenses, only 1115, 1520 and 1830 even give
$10\Gyr<\Delta\Tiso<20\Gyr$, let alone a consistent
$\Delta\Tiso=H_0^{-1}$.
\end{itemize}
Again we must keep in mind the caveats above, and also that the large
external shear in 1115, 0957 and 0911 means that for these lenses
$\Delta\Tiso$ properly speaking requires a modification given in WMK
but disregarded here. On the other hand, it seems unlikely that these
caveats will solve the serious discrepancies we see.  It appears more
likely that most real lenses do not belong to the generalized
isothermal family.

Whereas $\Delta\Tiso$ is rejected, are other scalings possible that
improve upon $\Delta T$?  L.L.R. Williams (personal communication)
points out that the definition (Equation \ref{sdel}) of $\Delta T$
considers the size of the lens but not its asymmetry, and that if we
multiply $(\theta_1+\theta_2)^2$ in the definition by a further factor
of $\sqrt{(\theta_1-\theta_2)/(\theta_1+\theta_2)}$ as a measure of
asymmetry, then the scaled time delays would range over a factor of
only 2.5, with no significant trend.  But the meaning of such an
asymmetry correction in terms of lensing theory is not known.

\section{Modelling the range of $\varphi$}

From the above, it appears that the scatter in $\varphi$ reflects a
range of mass profiles and source positions, and that its value must
be inferred for each lens by detailed modelling.  But without going
into detailed models for nine lenses, we can at least check whether
the observed range of $\varphi$ is plausible.

\begin{figure}
\centering
\includegraphics[width=.2\textwidth]{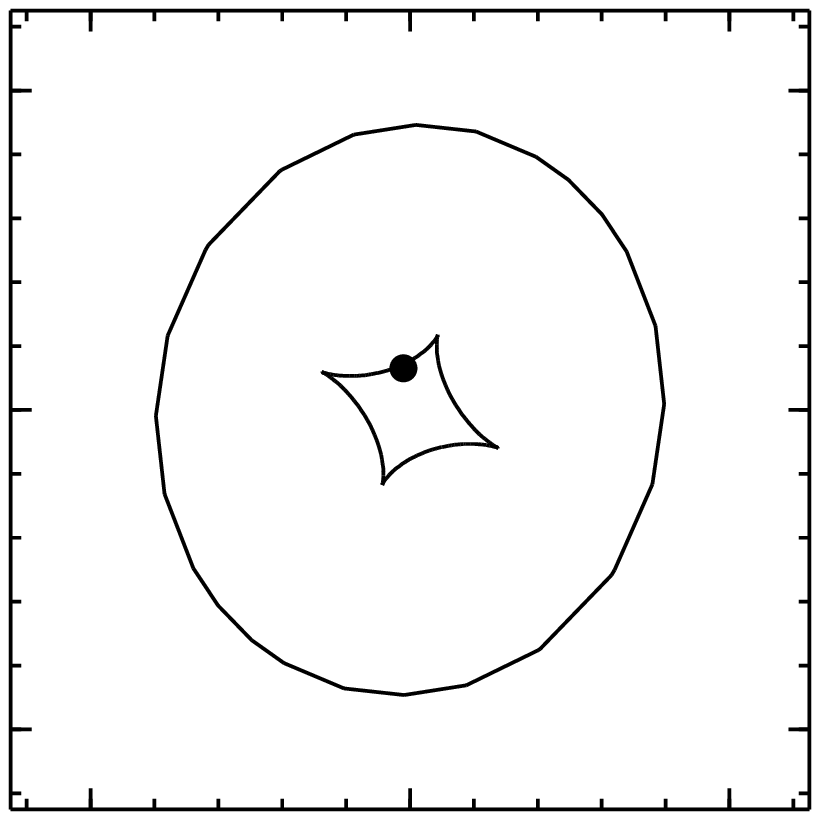}
\includegraphics[width=.2\textwidth]{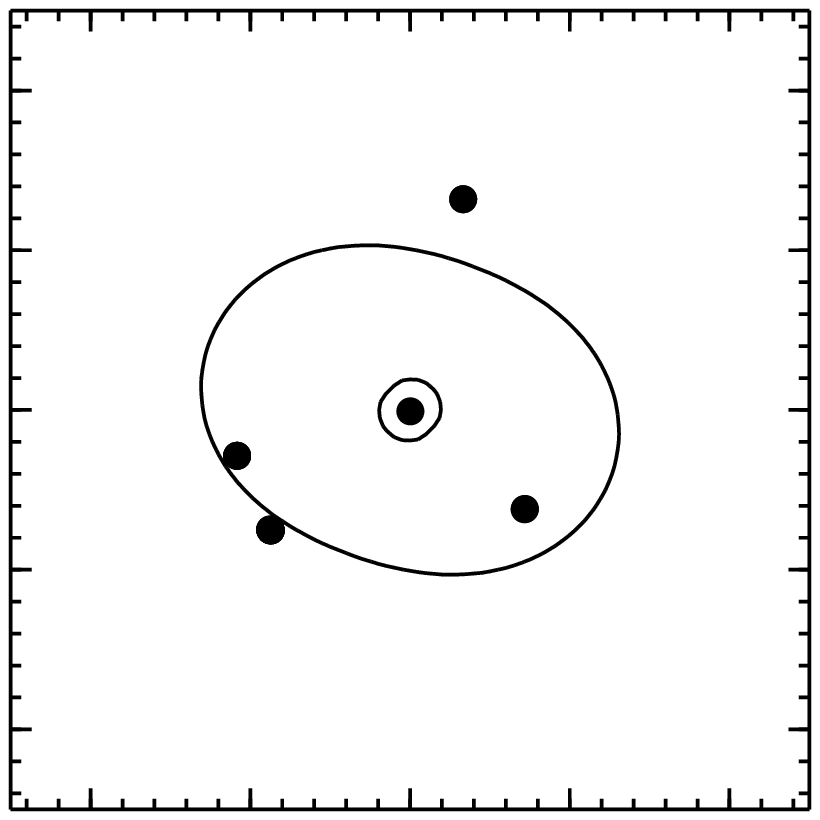}
\includegraphics[width=.4\textwidth]{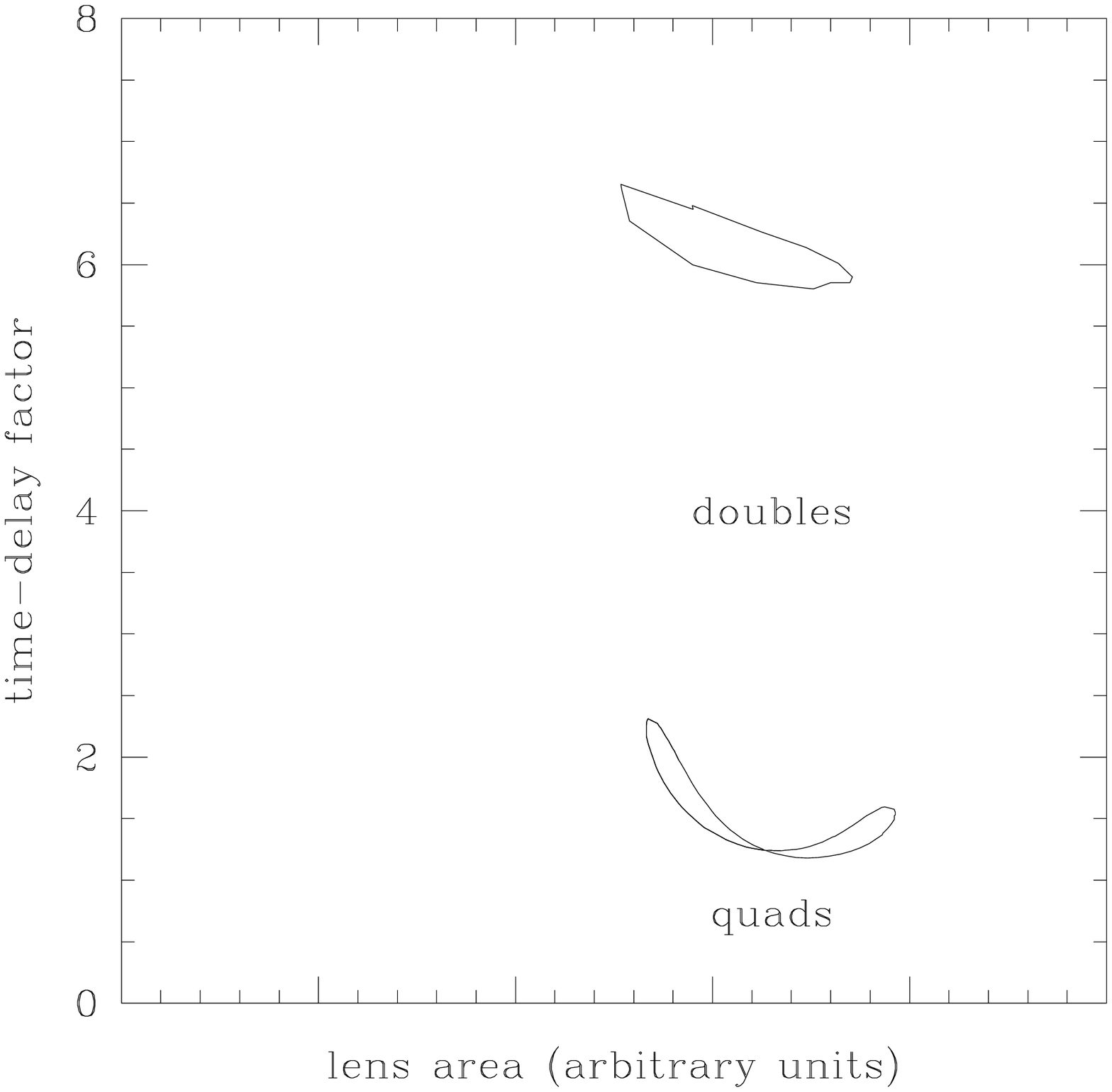}
\caption{Computation of $\varphi$ values from a simple model of
1115+080, taken from \cite{sw03}.  The top two panels show an image
morphology similar to 1115+080, and the corresponding source position.
The lower panel shows $\varphi$ against $(\theta_1+\theta_2)^2$ for
source positions along the two caustics.  [The horizontal axis is not
labelled because $(\theta_1+\theta_2)^2$ has arbitrary units:
arcsec$^2$, steradians, etc.]  The lower loop corresponds to the
diamond caustic and the upper loop corresponds to the outer
caustic. Hence quads are below the lower loop and doubles are between
the two loops.}
\label{fig-pg}
\end{figure}

Figure~\ref{fig-pg} shows such a check. The main plot is of $\varphi$
against the area $(\theta_1+\theta_2)^2$ for an example model (an
elliptical isothermal potential plus external shear.) The value of
$\varphi$ is shown for different source positions, the two loops
corresponding to source positions along the two caustics (actually
just inside the caustics, to avoid computational problems).  Quads are
below the lower loop, with $\varphi\leqsim2$. Doubles are between the
two loops, with $2\leqsim\varphi\leqsim6$.\footnote{Note that Figure
\ref{fig-pg} does not show a probability distribution, unlike related
plots in \cite{oguri02}.  The aim in Figure \ref{fig-pg} is simply to
show the separation of $\varphi$ for quads and doubles.}  The values
are model-dependent---for example, a steeper model will have both
loops somewhat higher.  Also, the value of $(\theta_1+\theta_2)^2$
depends on the source position: smaller for sources along the long
axis of the potential, larger for sources perpendicular to that axis.
But with these qualifications, Figure~\ref{fig-pg} shows that the
general ranges of $\varphi$, including the separation of quads and
doubles, is just as it is in the data, and there is no evidence that
the observed systems come from drastically different populations of
lenses.

\section{Summary}

We see in this paper a new interpretation of lensing time delays:
$\Delta t$ is $H_0^{-1}$ shrunk by the lens's covering factor on the
sky, times a number of the order of unity.  On separating off a redshift
dependent-term (also of order unity) we are left with a number
$\varphi$ (say) that summarizes the dependence on details of the lens
and lens configuration.

Using these ideas, we can rescale the observed time delays for the
nine currently-measured systems.  The observed time delays range over
a factor of 40, but the rescaled delays range over a factor of 5.  The
latter is the inferred range of $\varphi$, and moreover it appears
that $\varphi\leqsim2$ for quads and $2\leqsim\varphi\leqsim6$.
Reassuringly, the same spread in $\varphi$ is reproduced by a simple
model.

Using rescaled time-delays we can also test the hypothesis that the
observed lenses all belong to a generalized-isothermal family.  This
hypothesis is ruled out: it over-predicts time delays for large
lenses.  On the other hand, there is no indication that the known
time-delay systems come from drastically different types of lenses.

\appendix

\section{Significance of trends}

In Figures~\ref{fig-skyf} to \ref{fig-wmk} we have some points
$(x_i,y_i)$ and we want to know whether there is any trend in the
scatter.  There are many statistical tests relating to the
significance of trends in data, but none of the standard ones address
quite this question.  However, it is not difficult to design a
suitable statistical test.  Let us pose the question: what is the
probability of improving the fit to $y=\rm constant$ by shuffling the
$y_i$?  If nearly all shufflings reduce the $|\rm slope|$ we would
conclude that the data have a trend.

In the familiar straight-line fit, the slope is monotonic in
$\sum_ix_iy_i$.  Hence as a statistic, $\sum_ix_iy_i$ is equivalent to
the slope.

In the main text, I use the phrase ``significant at the 95\% level''
to mean that 5\% of shufflings increase the $|\rm slope|$.
Statisticians might use a phrase like ``$p$-value of 95\%''.

\begin{acknowledgements}
I am grateful to Rodrigo Ibata and Liliya Williams, who contributed
some fruitful suggestions.
\end{acknowledgements}

\end{document}